\documentclass[preprint2]{aastex}

\newcommand{\pt}{\partial}

\newcommand{\beq}{\begin{equation}}
\newcommand{\eeq}{\end{equation}}
\newcommand{\bea}{\begin{array}}
\newcommand{\eea}{\end{array}}

\slugcomment{To appear in ApJ}
\shorttitle{Stability of protoplanet systems}
 \shortauthors{ZHOU, LIN, SUN}

\begin{document}

\title{Post-Oligarchic Evolution of  Protoplanetary Embryos and the Stability of
      Planetary Systems}

\author{Ji-Lin Zhou\altaffilmark{1},  Douglas N.C. Lin\altaffilmark{2,3}, Yi-Sui Sun\altaffilmark{1}}
\altaffiltext{1}{Department of Astronomy, Nanjing University,
Nanjing 210093, China; zhoujl@nju.edu.cn; sunys@nju.edu.cn}
\altaffiltext{2}{UCO/Lick Observatory, University of California,
Santa Cruz, CA 95064, USA; lin@ucolick.org}
\altaffiltext{3}{KIAA, Peking University, Beijing 100871, China}

\begin{abstract}
  In the sequential accretion model, planets form through the
sedimentation of dust, cohesive collisions of planetesimals, and
coagulation of protoplanetary embryos prior to the onset of
efficient gas accretion.  As progenitors of terrestrial planets and
the cores of gas giant planets, embryos   have comparable
masses and are separated by the full width of their feeding zones after the oligarchic growth.
In this context, we investigate  the orbit-crossing time ($T_{\rm c}$) of
protoplanet systems both with and without a gas-disk background.
The protoplanets are initially with equal masses and
separation (EMS systems) scaled by their mutual Hill's radii. In a gas-free environment,
 $\log (T_{\rm c}/{\rm yr})\simeq A+B \log
(k_0/2.3) $, where $k_0$ is the  initial separation of the
protoplanets normalized by their Hill's radii, $A$ and $B$ are
functions of their masses and initial eccentricities. Through a
simple analytical approach, we demonstrate that the evolution of
the velocity dispersion in an EMS system follows a random walk.
 The stochastic
nature of random-walk diffusion leads to (i) an increasing average
eccentricity $<e> \propto t^{1/2}$, where $t$ is the time; (ii) Rayleigh-distributed
eccentricities ($P(e,t)=e/\sigma^2
 \exp(-e^2/(2\sigma^2))$, where $P$ is the probability and $\sigma(t)$ is the dispersion) of the protoplanets;
  (iii) a power-law dependence of $T_{\rm c}$ on
 planetary separation. As
evidence for the chaotic diffusion,  the
observed eccentricities of known extra solar planets can be
approximated by a Rayleigh distribution.
  In a gaseous environment,
 eccentricities of the protoplanetary embryos are damped by
their interactions with the gas disk on a time scale $T_{\rm
tidal}$ which is inversely proportional to the surface density of
the gas. When they become well separated (with $k_0 \simeq 6-12$),
the orbit-crossing tendency of embryos is suppressed by
the tidal drag and their growth is stalled along with
low-eccentricity orbits.
However, the efficiency of tidal damping
declines with the gas depletion.
We evaluate the
isolation masses of the embryos, which  determine the probability of gas giant
formation, as a function of the dust and gas surface
densities. Similar processes regulate the early evolution of
multiple gas giant planet systems.

\end{abstract}


\section{Introduction}

The origin and evolution of multiple planets around the Sun have been
the primary stimulus for the studies of classical N-body systems.
The pioneering analysis of Poincar{\'e} (1892) established the paradigm
that the motion of systems with more than two bodies is not integrable.
This fundamental result is well supported by the dynamical
diversity among the $\sim 200$ extra solar planetary systems discovered in the
past decade.  While considerable efforts have been made to interpret
various physical processes which may contribute to their new-found
properties, it may be fruitful to explore, in depth, the implication
of long-term dynamical interactions among members of multiple-planet
systems. These analysis may eventually provide the basis for a theory
of statistical mechanics which characterizes the architecture of planetary
systems.

Statistical mechanics has been employed to study other N-body systems in
astrophysics.  In the context of stellar clusters,
the time scale of phase-space relaxation may be evaluated by a Fokker-Planck
approximation.  The magnitude of the diffusion coefficient is determined
by an impulse approximation, {\it i.e.} as an ensemble of independent
close encounters.  But in planetary systems, the host stars dominate the
gravity field. Although planetary perturbations are weak, they
persist and are correlated over many orbits.  This aspect of the dynamical
evolution makes the development of a statistical approach particularly
difficult.

The investigation of the phase space diffusion  is closely related to
the stability of planetary systems.
 With the exception of periodic and quasi-periodic orbits,
  the stability of most orbits in a general
N-body planetary system is not known. The Kolmogorov-Arnold-Moser
(KAM) theory proved that a non-degenerate integrable Hamiltonian system may
preserve most of its stable (quasi-periodic) motions under sufficiently small
and analytical perturbations \citep{kol54,mos58,arn63}. For those non-stable
motions, the Nekhoroshev theorem  showed that, the time that an orbit becomes
unstable grows exponentially with respect to the inverse of the non-integrable
parameter \citep{nek77}.  For vanishing ``perturbation'' amplitude, the
diffusion time scale become infinitely long. However, most systems of
astronomical interest, such as planetary systems, are degenerate. Consequently,
the applications of the powerful KAM and Nekhoroshev theorems turned out to be
indirect and difficult \citep{sm71,mg97}.

Nevertheless, the stability of planetary systems remains an important problem
with many applications. The first application of this fundamental issue concerns the
dynamical age of the Solar System. Although interactions between the planets
give rise to chaotic motions, the system is expected to remain essentially
stable over a time much longer than its present age of 4.6 Gyr \citep{las89,sw92,mh99}.

Another issue is the stability of a proto-planet system during the early stage
of its formation. According to the conventional sequential-accretion scenario,
the terrestrial planets are formed by the coagulation of planetesimals in
protostellar disks \cite{saf69, wet80}. Through several stages of
runaway and oligarchic growth, cohesive collisions lead to the emergence of
massive protoplanetary embryos \cite{ki02,il04}. According to the numerical
simulations  \citep{ki98}, protoplanets form with comparable masses and similar
separation ($\sim 10$ Hill's radii). The stability of such protoplanet systems
could be crucial for the subsequent evolutions and final configurations of the
system, like the presence of  Earth-mass planets near their host stars (e.g., Zhou et al. 2005).

A third issue concerns the excitation of the large  eccentricities as well as the
stability of the recently observed extra solar planet systems\footnote{http://exoplanets.org/,
http://vo.obspm.fr/exoplanetes/.}. The observed extra solar planet systems
have a median  eccentricity of $0.25$ \cite{mar05}.  Despite its large uncertainties,
the eccentricity distribution of extra solar planets is quite different from
our Solar System. As interactions between gaseous disks and protoplanets are expected to
generally limit their eccentricities \cite{pap06}, the origin of
the large eccentricities in extra solar systems remains poorly understood.

Despite these important questions, an analytic theory for stability of planetary systems
has not been attained. Facing this enormous complexity, recent attempts to understand some
aspects of this process have been reduced to a subset of three-body problems.
%Wisdom (1980) found that
%a 2-planet-plus-the-star system with equal planetary masses will be
%chaotic as long as the separation of the  two planets $< 2 \mu^{2/7}$,
%where $\mu$ is the mass ratio of each planet to the star.
 Based on the
results from qualitative studies of the general three-body problem (e.g., Marchal 1990),
Gladman (1993) investigated the stability of the two planet systems both
analytically and numerically. He found that a system of two planets  with mass ratios to
the star $\mu_1,\mu_2$ could be Hill stable if
their separation $> 2\sqrt{3}( \frac{\mu_1+\mu_2}{3})^{1/3}$, where
 Hill stable is defined as orbits that will never cross.
%Thus the  Hill unstable region is  smaller than the chaotic region
% provided $\mu_1=\mu_2=\mu \le 2\times 10^{-4}$.
 In systems with more than two planets,
the most practical approach is to resort to numerical simulations.  Due to the large degrees
of freedom of these systems, restrictions are needed to reduce the range of configurations
for parameter studies. Motivated by the characteristics of embryo systems after runaway
and oligarchic growth, a series of investigations have been carried out to study
idealized but well-defined planetary systems with equal masses and scaled separation.
Hereafter we refer these idealized planet systems as EMS systems.

Chambers et al. (1996) determined numerically the orbital crossing time $T_{\rm c}$ of EMS systems
with $n$ planets $(n\ge3)$ initially on circular orbits. %\cite{cwb96}.
They found an exponential relation $\log T_{\rm c} \sim k_0 $, which seems to be independent
of $n$. The dimensionless parameter $k_0$ is the scaled initial separation.  They did not provide any explanation of the underlining cause of this relation.
 Later, Yoshinaga, Kokubo and Makino (1999) generalized this study to the cases that the planets are initially on non-circular
and non-coplanar orbits.  In the limit of small initial eccentricity $e_0$ and inclination, they
obtained similar results as previous investigators. Later, the instability of EMS systems
under solar nebular gas drag was studied by Iwasaki  et al. (2001, 2002) and Iwasaki \&  Ohtsuki (2006).

  However, the EMS systems studied in these
works are with separation $k_0<10$.
 %and without any interactions between the protoplanets (or embryos) and the gas disk.
For realistic planetary systems, the initial separation between
planets may be larger, with a gas disk during the stage of planet formation.
In the Solar System, the present-day values of $k_0 \sim 8-64$.
According to the numerical simulations of planet formation (Kokubo \& Ida 2002, Ida \& Lin 2004),
after the planetary embryos have depleted nearby planetesimals
and reached  isolation masses, the embryos were separated with $k_0\sim 10-12$.

 The initial motivation of the present work is to extend the previous studies to
the cases $k_0> 10$ both with and without a gas disk, and to derive
a functional dependence of $T_c$ on $k_0,\mu, e_0$.
We show in \S2 that, the orbit crossing time $T_{\rm c}$ is better approximated by a power-law relation
$\log T_{\rm c} \sim \log k_0$. A simple analytical interpretation of this relation is suggested in \S3.
We also show that the average eccentricity of an EMS system in a gas-free environment increases
as $\sim t^{1/2}$.  We identify this evolution as a result of the random walk
diffusion in phase space which accounts for the power-law dependence of the orbital
crossing time on the initial separation. In \S4, we extend the study to the cases when the
protoplanets (or embryos) are embedded in a gas environment.  This investigation determines
the range of feeding zones and isolation masses of embryos in gas-rich protostellar disks.
The embryos' masses and separations during the post-oligarchic evolution in
a depleting gas environment are derived.
  These quantities determine the probability of gas giant formation.  We show that
the observed eccentricity distribution of known extra solar planets has the form of
a Rayleigh distribution.  We cite this property as evidence for chaotic diffusion being
the dominant excitation mechanism. Summary and the implications of our results on the
formation of planet systems are presented in the final section.

\section{Empirical formula for $T_{\rm c}$ without gas disk}

The model of an EMS system is given as follows.  Suppose $n$ protoplanets
(or planets for simplicity) with equal masses move around a star with one solar
mass, and the separation between them are equal when scaled
by their mutual Hill's radii. In this paper all the orbits of the planets are
 coplanar, especially the EMS systems are in a gas-free environment
 in this and the coming sections.

We denote the mass ratios of the planets to the star,
 the semi-major axes and eccentricities of the
planets' orbits as $\mu$, $a_i$ and $e_i$ (i=1,...,n), respectively. The
scaled  separation and eccentricities of the planet orbits are
\beq
\begin{array}{l}
 k =  \frac{a_{i+1}-a_i}{R_H}, ~(i=1,...,n-1), \\
 \tilde{e_i}=\frac{e_{i}}{h}, ~~ (i=1,...,n),
 \end{array}
\label{kbe} \eeq
respectively, where $R_H$ is the mutual Hill's radius and   $h$ is the relative separation of two neighboring
planets, defined as
 \beq
R_H= (\frac{2\mu}{3} )^{1/3}
\frac{a_i+a_{i+1}}{2}, ~h=\frac{a_{i+1}-a_i}{a_{i+1}+a_i}.
\label{eq2}
 \eeq
 Thus the orbits of two neighboring planets with
$\tilde{e}=1$ will cross if the difference between their perihelion angles  is $\pi$.
For simplicity, we adopt the same initial eccentricities
$\tilde{e}_0$, while the initial mean anomaly $M_i, (i=1,...,n$), and longitude
of perihelion $\varpi_i$ of each planet are chosen randomly.
We take $n=9$, and arbitrarily specify the initial semi-major axis of the fourth planet
  $a_4=1$AU for normalization purposes. So when the initial separation $k_0=k(t=0)$ varies,
  the planet system is enlarged both inward and outward.

\begin{figure}
 \vspace{5cm}
\includegraphics{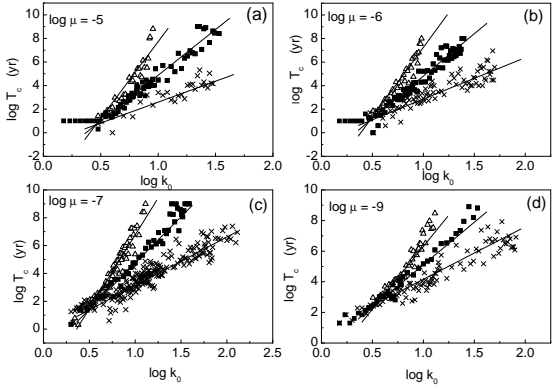}
 \caption{\small Variations of the orbit-crossing  time $T_{\rm c}$
with initial orbital separation $k_0$ in the 9-planet EMS systems of
different $\mu$ and $\tilde{e}_0$. The triangles, squares and
crosses denote systems with $\tilde{e}_0=0, 0.5, 0.9$, respectively.
 The solid lines are calculated from the empirical formula (\ref{ff}).
 In the $\mu=10^{-9}$  case (d),  a correction of $+0.5$ is added to the
 values of $\log ~T_{\rm c} $  given by  equation (\ref{ff}).
 \label{fig1}}
\end{figure}

 The orbital crossing time of the EMS system (denoted as $T_{c}$) is defined as
 the minimum duration  when either of the following two situations occurs
 between any two planets during the evolution:
(1) close encounter, defined as the distance between them is
less than their mutual Hill's radius, (2) orbit crossing, defined
as $a_i \ge  a_{i+1},(i=1,...,n-1)$. We use the symplectic code of Wisdom and
Holman (1991) from the SWIFT package (Levison \& Duncan 1994).
Whenever orbit crossing or a close encounter occurs, we halt
the integration. The time step is chosen to accommodate $\sim 20$
steps per inner planet orbit, and the  accumulated error of the relative
energy during the integration is constrained to be $\sim 10^{-10}-10^{-9}$ until the
system becomes unstable.
% Fig.1 shows the evolution of energy error of
%a system  with crossing time scale $>10^9$ yr.

We investigate mainly 7 typical values of
$\mu=10^{i},(i=-10,...,-4)$. For each value of $\mu$, we do 10
sets of simulations with initial eccentricities of the
planets in the range $\tilde{e}=0,0.1,0.2,...,0.9$. For each
set of parameters, many orbits
with various initial value $k_0 $ are integrated
to determine the relationship between $T_{\rm c}$ and $k_0$.

Fig.1 shows the dependence of $T_{\rm c}$ on $k_0$ for a range of $\mu$.
We find there exists roughly a critical $k_{\rm c}$ such that,
$T_{\rm c}$ is independent of $k_0$ for $k_0<k_{\rm c}$ and
increases with $k_0$ for $k_0>k_{\rm c}$(Fig.1a,1b). These two branches of solutions join continuously at $k_0=k_{\rm c}$
with the approximation $T_{\rm c}(k_0 = k_{\rm c}) =A$. We are primarily
interested in the range of $k_0>k_{\rm c}$ for which the numerical results
can be fitted with $ \log (T_{\rm c}/\rm yr) =A + B \log(k_0/k_{\rm c})$.
 In order to obtain the value of
the numerical coefficients, $A$, $B$, and $k_{\rm c}$, we proceed as follows:
\begin{description}
\item[(i)]We first determine $k_{\rm c}$ by scaling $T_{\rm c}$ with $k_0$ in the range
$[1.5,3.5]$. We found the eccentricity-dependence of $k_{\rm c}$ to be negligible
over $\tilde{e}\in [0,0.5]$. For the entire range of $\mu$, we obtain
$k_{\rm c} \approx 2.3$, again insensitive to the magnitude of $\mu$ (Fig.2a).
\item[(ii)]We evaluate the  average values of $A=T_{\rm c}(k_0 =k_{\rm c})$, and find
$A=(-0.91\pm 0.08)-(0.27\pm 0.01) \log \mu$ (Fig.2b). A more general expression,
which  also incorporates the eccentricity dependence of $T_{\rm c}$, is
$A=-2+{\tilde e}_0-0.27\log \mu$.
\item[(iii)]Finally, we determine the magnitude of B. From the slopes of the
$\log(T_{\rm c})-\log(k_0)$ curves of Fig.1, we obtain the eccentricity and $\mu$
dependence of $B$ (Fig.2c-d). A reasonable approximation for the $B
(\mu, {\tilde e}_0)$ is $ B=
   b_1+b_2\log\mu+(b_3+b_4 \log\mu) {\tilde e}_0$,
   with  $b_1=18.7\pm 0.6, ~~b_2=1.11\pm
  0.08, ~~b_3=-16.8\pm 0.6, ~~b_4=-1.24\pm 0.08$.
\end{description}
After some exhaustive simulations, we obtain the following empirical fitting formula:
\beq
\begin{array}{l}
 \log (\frac{T_{\rm c}}{\rm yr}) =A + B \log(\frac{k_0}{2.3}).  \\
 (k_0>2.3, 10^{-4}\le \mu \le 10^{-10})
 \end{array}
 \label{ff}
\eeq
where
\beq
\begin{array}{l}
A=(-2+\tilde{e}_0-0.27\log\mu) \\
B=(18.7+ 1.1\log\mu)-(16.8+1.2\log\mu ){\tilde{e}}_0.
\end{array}
\label{ab}
\eeq

\begin{figure}
 \vspace{5cm}
\includegraphics{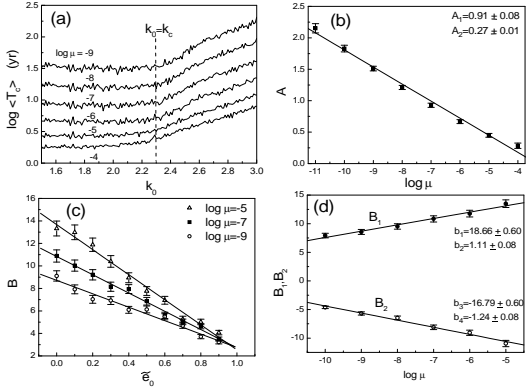} \caption{\small The procedure to determine the coefficients
$k_c, A,B$ in formula (\ref{ff}). (a) Variations of the average
$T_{\rm c}$  with small $k_0$. The
  average is taken over $\tilde{e}\in [0,0.5]$. From bottom to up,
  the curves correspond to EMS systems with $\mu=10^{-4},...,10^{-9}$,respectively.
  $k_{\rm c}$ is defined so that $<T_{\rm c}>$ begins to increase with $k_0$ at
  $k_0>k_{\rm c}$.
    (b) Determine $A=<T_{\rm c}>(k=k_c)$ for  different $\mu$. The squares with error bars are
    numerical results, while the solid line ($A=A_1+A_2 \log\mu$)
    is the best-fit line.  The best-fit coefficients are also shown.
  (c)  The triangles, squares and circles with error bars denote the best-fit
     slopes $B$ of the curves ($\log(T_{\rm c})-\log(k_0)$) in Fig.1. As a function of
$\tilde{e}_0$,  it can be expressed as  $B=B_1+B_2 {\tilde{e}_0}$
for various $\mu$. The  best-fit coefficients for $B_1= b_1+b_2 \log
(\mu)$ and  $B_2= b_3+b_4 \log (\mu)$ are shown in (d).
  \label{fig2}}
\end{figure}

The predictions given by the formula (\ref{ff}) are plotted also
in Fig. 1. We find the formula agrees well with the numerical
results for planetary masses $10^{-4}\le \mu \le 10^{-10}$.  In
this mass range, slope $B$ is positive. The above formula
(\ref{ff}) generalizes a similar approach introduced by Chambers
et al. (1996)\footnote{For $\tilde e_0 =0$ and $\mu=10^{-7}$,
Chambers et al. (1996) found $ \log T_{\rm c}= b k_0+c$ in the
range $k_0< 10$, with $b=0.76\pm 0.03 $ and $c=-0.36 \pm 0.18 $.
They also obtained similar expressions for other values of $\mu$.
This expression can be obtained from equation (\ref{ff}) in the limit
of small $k_0$.  For example, in the range of $k<10$, $x \equiv
(k_0-6)/6 <1 $ and equation (\ref{ff}) reduces to $\log T_{\rm c}  =
11[ \log (1+x)+\log(\frac{6}{2.3})]-0.11\approx \frac{11}{\ln 10}x
+4.47 = 0.80 k_0 -0.31$.}. The distribution of $T_{\rm c}$ in the
separation-mass ($k_0-\mu$) space is shown in Fig. 3a for ${\tilde
e}_0=0$.

 However,
we find formula (\ref{ff}) is not satisfied when applied to $\mu\sim 10^{-3}$.
Since in these situations, resonances between planets are strong and dominate the
dynamics at the place $k_0=2(\frac{q-1}{q+1})/(\frac23\mu)^{1/3}$, where
$q=(n_{i}/n_{i+1})^{2/3}$ is the ratio of the mean motions of planets $i$ and $i+1$.
 As $\mu\sim 10^{-3}$
is the ideal case for giant planet systems, we investigate this case for planets on
initial circular orbits, and find the orbital crossing time can be
approximated by a simple formula in the case $k_0 < 10$:
\beq
    \log (\frac{T_{\rm c}}{\rm yr})\approx -5.0+2.2 k_0. ~(\mu\sim 10^{-3},\tilde{e}=0)
\label{f3m}
\eeq
Fig.3b shows the numerically determined orbital crossing time with the best fit formula
(\ref{f3m}). The drop of $T_{\rm c}$ near $k_0\sim 5$ is due to the presence of the $2:1$ resonance
($k_0 \simeq 5.2$) between the planets.

\begin{figure}
\vspace{3cm}
\includegraphics{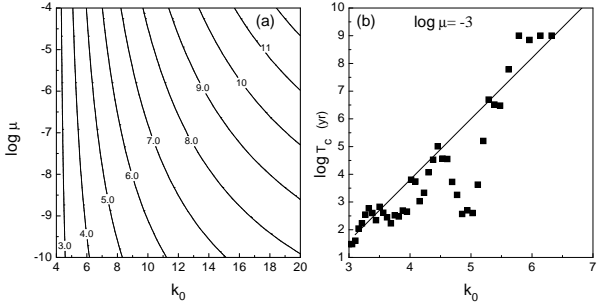}
\caption{\small The orbital crossing time on
parameter space. (a): Contour lines of
  $\log(T_{\rm c})$ of EMS systems in circular orbits
 in the space of initial orbital separation $k_0$ and
 planet masses $\mu $. The numbers in the curves are $\log(T_{\rm c})$.
They are obtained from formula (\ref{ff}). (b): Variations of $T_c$
on $k_0$ for  $\mu=10^{-3}$. Squares are from numerical simulations,
and the solid line is from formula (\ref{f3m}). The big drop at
$k_0\sim 5$ corresponds to $2:1$ resonance between planets.
\label{fig3}}
\end{figure}

 From equation (\ref{ff}), we can highlight the difference in the crossing time of two EMS systems
(denoted as S1 and S2,respectively) on initial circular orbits:

\begin{itemize}
\item Suppose S1 and S2 have the same planetary masses: $\mu_1=\mu_2=\mu$,
 \beq
 \frac{T_{c1}}{T_{c2}}=(\frac{k_{01}}{k_{02}})^{18.7 + 1.1 \log \mu}.
 \eeq
 Thus for example, if $\mu=-7$ and $k_{01}/k_{02}=2$, the above formula
 yields $T_{c1}/T_{c2}\approx 2000$. The
 crossing time of the widely separated system (S1) is three orders of magnitude larger than that of the compact
 system (S2), even though the initial separation among planets  differs only by a factor of 2.
\item In contrast, let S1 and S2 have the same planet separation $k_{01}=k_{02}=k_{0}$,
   \beq
   \frac{T_{c1}}{T_{c2}}=(\frac{\mu_{1}}{\mu_{2}})^{-0.27 +1.1 \log(k_0/2.3)}.
   \eeq
Thus for example, if $k_0=10$ and $\mu_{1}/\mu_{2}=10$,
it gives  $T_{c1}/T_{c2}\approx 2.7$.
The crossing time for the massive system (S1) is around three times longer than the
 less massive system
(S2), provided their normalized (by the Hill's radius) separations are the same.
\end{itemize}

\section{A simple  analytical approximation}
The numerical simulations, though informative,
 do not provide any underlying theory for the origin of
the  dependence of $T_{\rm c}$ on $k_0$, $\mu$ and $\tilde{e}_0$.
 In this section, we present a simple
analytical approach in an attempt to describe the evolution of the EMS systems without gas disk.
We identify the planets of an EMS system with subscript $l$ ($1,2,...,l-1,l,l+1,...,n$
with $n\ge 3$), in the increasing order of their initial semi-major axes.
We consider the evolution
of a representative planet $1<l<n$.
Assume all the planets are initially on circular orbits,
and in the limit of close separation, {\it i.e.} $a_{l+1} - a_l < < a_l$.
According to equations (\ref{kbe}) and (\ref{eq2}), this approximation is
equivalent to   $k_0(2\mu/3)^{1/3}\ll 1$. We call it the close separation
assumption.
The largest perturbations on planet $l$
come from close encounters with nearby planets (planet $l\pm 1$).
%, we assume the total response of
%planet $l$ is a linear combination of the perturbation by its two neighbors.
 Under the close separation assumption, the interactions between
each pair of neighbors can be well approximated by an independent set of Hill's problems.

\begin{figure}
 \vspace{3cm}
\includegraphics{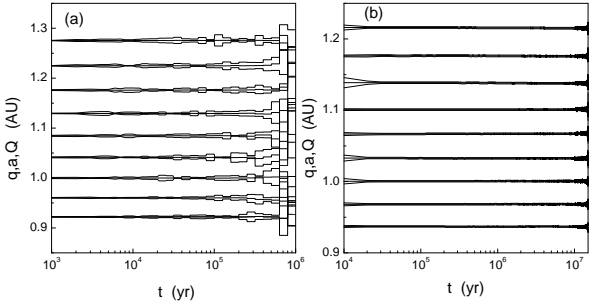} \caption{\small  Evolution of $q=a(1-e),a,Q=a(1+e)$
for the 9-planet EMS system in a (a) gas-free, (b) gas-rich
environment. Parameters in (a) are $\mu=10^{-7}$, $e_0=0$, $k_0=8$.
 The orbital crossing time is $7\times 10^5 $ yr, according to equation (\ref{ff}).
 Parameters in (b) are $\mu=10^{-7}$, $e_0=0.5h$, $k_0=8$.
 The orbital crossing time is $1.5\times 10^7 $yr.
 From Fig.3 and formula (\ref{ff}), the orbital crossing time for
 the same parameters but in a gas-free environment is $\sim 10^4$ yr.
\label{fig4}}
\end{figure}

We define $\epsilon \equiv (a_l-a_{l-1})/a_l \simeq k_0 (2\mu/3)^{1/3}$
as the relative semi-major axis, $z_l \equiv e_l\exp( i\varpi_l)$ as the Runge-Lenz vector,
and $\varpi_l$ as the longitude of periapse of planet $l$. We consider the limit $e_l \ll \epsilon \ll 1$.
To first order in $\mu$, $a_l,a_{l-1}$ do not change during close encounters (H\'enon
\& Petit 1986). We assume that during all close encounters prior to orbit crossing the
semi-major axes of the planets do not have significant secular changes.  This assumption
is supported by the numerical results (See Fig.4a).
However, $z$ evolves and after the $j$-th close counter between the planets $l-1$ and $l$,
the change in $z$ is given as %\citep{hp86},\citep{dun89}
\beq
z_{j}=z_{j-1} - i \frac{g \mu}{\epsilon^2}   \exp(i \lambda_{j-1}), ~(j\ge 1),
 \label{zj}
 \eeq
where $\lambda_j$ is the mean longitude of planet $l$ ,  $g=\frac89 [2K_0(\frac23)+K_1(\frac23)]
\approx 2.24$, where $K_0$ and $K_1$ are modified Bessel functions (H{\'e}non \& Petit 1986,
Duncan, Quinn \& Tremaine 1989). The time between two consecutive close encounters is given as
$T_s=T_l [(a_{l}/a_{l-1})^{3/2}-1]^{-1} \approx \frac23 T_l \epsilon^{-1} $,
where  $T_l$ is the orbital period of the planet $l$.

For illustrative purposes, we adopt $a_l=1$ AU, so $T_l=1$ yr,
and the change of $\lambda$ during one encounter is given as
$\lambda_{j}\approx \lambda_{j-1}+\frac{4\pi}{3 \epsilon }
\label{lj} $ . Since $\epsilon \ll 1$
and the change of $\epsilon$ is  second order
in $\mu$, $\lambda_j~(j=1,...,n)$ at successive encounters behave
like a series of random numbers in $[0,2\pi]$.
According to (\ref{zj}) we have,
\beq
e^2_{j}-e^2_{j-1}=-2 \frac{ g \mu }{\epsilon^2}e_{j-1}
\sin(\lambda_{j-1}-\varpi_{j-1})+\frac{g^2 \mu^2 }{\epsilon^4}.
\label{sig}
\eeq
 Due to the near-random phase of $\lambda_{j}$, the first term in equation (\ref{sig}) averages
to zero over a long time. Changes of $e^2$
induced by the perturbations from planets $l\pm 2,l\pm 3,...$  are
  $\sim 1/2^4,1/3^4,...$ times those
from $l\pm 1$.  However, the periods of close
encounters between planet $l$ and these planets are
$\sim 1/2,1/3,...,$ times $T_{\rm s}$, respectively.
 Therefore, when we take account of perturbations from more distant
planets on both sides, we introduce a factor $2(1+1/2^3+1/3^3+...)\approx 2.40$,
% Under the assumption that the total response of planet $l$ is the linear
%combination of perturbation due to both closest neighbors, which is justified
%by the random phase of $\lambda_{j}$, we double the right hand
% side of eq. [\ref{sig}] to obtain
so that $ <\Delta e^2> = 2.4 g^2 \mu^2 \epsilon^{-4}$.
The average eccentricity of the $l$-th planet after $j$ close encounters
with nearby planets is estimated to be
\beq
<e^2>^{1/2} = \sqrt{2.4} g\mu \epsilon^{-2}j^{1/2} \approx
5.2 k_0^{-3/2}\mu^{1/2} (\frac{t}{\rm yr})^{1/2},
\label{etf}
\eeq
where we have substituted $j=t/T_s= \frac{3}{2} \epsilon t/{\rm yr}$.
This formula will be confirmed by numerical simulations in this section.

According to the criteria specified in \S2, orbit crossing occurs when
$<e^2>^{1/2}\sim h= \frac12 k_0  (\frac23 \mu)^{1/3}$.  From equation (\ref{etf}),
we derive,
\beq
\log(\frac{T_{\rm c}}{\rm yr}) \approx -1.1 +5 \log k_0-\frac13 \log\mu.
\label{thf}
\eeq
This expression describes the power law dependence of $T_{\rm c}$ on $k_0$ as in equation (\ref{ff}).
However, the discrepancy between the coefficients
$B$ and $5$ in equations (\ref{ff}) and (\ref{thf})
 is considerable, especially when
$\mu$ is large. This may be due to the close separation assumption,
  $\epsilon  \sim k_0 \mu^{1/3} \ll 1$ no longer being valid
 for moderate $k_0$ and $\mu> 10^{-5}$.
  Moreover, the sign of the coefficient  of $\log \mu $ is negative which disagrees with equation (\ref{ff}).
 This may be caused by the oversimplified assumptions in the analytical model.

Next, we show that the evolution of the average eccentricity ($<e^2>^{1/2}
\propto t^{1/2}$) is mainly driven by a random walk process.  The stochastic
nature of the perturbations also leads to the power law dependence of $T_{\rm c}$ on
$k_0$. We define the velocity dispersion as $v \equiv |{\bf v}_{\rm kep}|-
|{\bf v}_{\rm cir}|$, where ${\bf v}_{\rm kep},{\bf v}_{\rm cir}$ are the
velocities of Keplerian and circular motion respectively. It is easy to
show that $v= na e\cos f+o(e^2)$, where $f$ is the true anomaly.
We consider  a group of orbits in phase space,
 and the probability of planet $l$ having velocity dispersion $v$ is
 denoted by $P(v)$. Thus $P(v)$ describes the
 distribution of a group of orbits in velocity dispersion
 space. Since every close encounter between planets
 will modify the distribution,
   $P(v)$ is a function of time $t$ (or $j$ encounters).
We assume that the planetary motions are chaotic and occupy
 a stochastic region in the phase space. This assumption is justified
by the random phase of $\lambda$ and the non-zero Lyapunov exponents
shown at the end of this section.

Under the chaotic assumption, the evolution of
$P(v,j)$ obeys the Fokker-Planck equation (Lichtenberg
\& Lieberman 1990):
\beq
\frac{\pt P}{\pt j}=-\frac{\pt }{\pt v}(BP)+\frac12 \frac{\pt^2}{\pt v^2}(DP),
\label{FK}
\eeq
where $B,D$ are the frictional and diffusion coefficients, respectively, with
\beq
\begin{array}{ll}
D(v) & =\frac{1}{2\pi} \int_{0}^{2\pi} [\Delta v(\psi )]^2 d\psi \\
  & =n^2a^2 \frac{1}{2\pi} \int_{0}^{2\pi} [\Delta e (\psi ) \cos f]^2 d\psi,
  \end{array}
\eeq
where $\psi=\lambda-\varpi$.  Following the  standard procedure in celestial mechanics,
we carry out orbit averaging around the Keplerian motion so that
$\cos ^2 f=1/2+o(e^2)$.  We adopt the approximation
$(\Delta e)^2 \approx \Delta e^2$. According to equation (\ref{sig}),
we find $D(v)\approx n^2 a^2 \mu^2 g^2 \epsilon^{-4}$.
Since $D$ is independent of $v$, $B=\frac12 \frac{dD}{dv}=0$.
After replacing $j$ by $t$, the Fokker-Planck equation   is converted
into the standard diffusion equation:
\beq
\frac{\pt P}{\pt t}=\tilde{D} \frac{\pt^2 P}{\pt v^2},
\eeq
where $ \tilde{D}=\frac34 \epsilon  D {\rm yr}^{-1} \approx 5.6 n^2a^2 \mu k_0^{-3}{\rm yr}^{-1}$.

The time dependent solution of the above equation with the initial value $P(v, 0) = \delta(0)$
(where $\delta(x)$ is the Dirac delta function) is a Gaussian (i.e., normal) distribution:
\beq
    P(v,t)= \frac{1}{\sigma\sqrt{2\pi}}\exp(-\frac{v^2}{2\sigma^2}), ~~ \sigma= (2 \tilde{D} t)^{1/2}.
\label{gau}
\eeq
Substituting $\tilde{D}$, we find
\beq
\frac{\sigma}{na}\approx 3.4 k_0^{-3/2} \mu^{1/2} (\frac{t}{\rm yr})^{1/2}.
\label{stf}
\eeq
We convert equation (\ref{gau}) to a distribution of
eccentricity by substituting $v=nae\cos f$,
where functions of $\cos f$ are replaced by the average values over a Keplerian period,
$<\cos f>=-e$ and $<\cos^2 f>=1/2$. Thus we get,
\beq
  P(e,t)=\frac{e}{\tilde{\sigma}^2} \exp(-\frac{e^2}{2\tilde{\sigma}^2}),~~ \tilde{\sigma}=\frac{\sqrt{2}\sigma}{na},
  \label{ray}
\eeq
which has the form of a Rayleigh distribution.

\begin{figure}
 \vspace{4cm}
\includegraphics{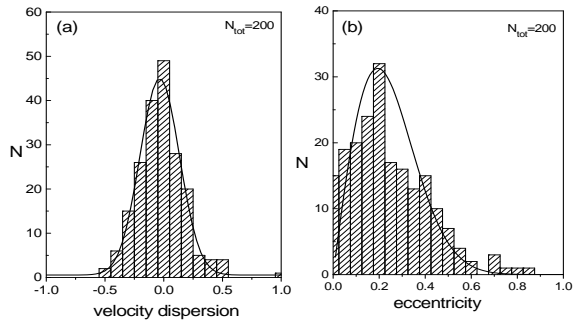}
\caption{\small Distributions  of (a) the velocity
dispersions $v$ and (b) eccentricities  in four runs of 50-planet EMS
systems with $\mu=10^5, k_0=5$ at time $t=0.4$ Myr. The fit
Gaussian distribution in (a) is according to equation (\ref{gau})  with $\sigma=0.336$,
an adjustment of  $<v>=-0.0342$, and  a scale factor of $37.4$.
 The fit of the Rayleigh distribution in (b) is according to  equation (\ref{ray}) with
 $\sigma=0.194$ and a scale factor  of $10$.%$<e>=0.243$,
 \label{fig5}}
\end{figure}

In order to verify the above analytical results, we carry out some numerical simulations
with EMS systems of $n=50$ protoplanets.  These results also provide a self-consistent verification
on the assumed chaotic nature of planetary motion.  In these simulations, we specify the following
initial conditions. The planets are initially placed on circular orbits, with $a_1=1AU$. We utilize
the Hermit scheme P(EC)$^3$ in order to follow the planets' evolution after their orbital
 crossing (Makino \& Aarseth 1992, Yoshinaga, Kokubo \& Makino 1999).
 %\cite{ma92};

Figs. 5 and 6 show some typical numerical results.  At each given epoch,
the normalized velocity dispersions relative to the circular orbits
follow a Gaussian distribution (\ref{gau}).  The corresponding eccentricities
obey a Rayleigh distribution (\ref{ray}) (see Fig.5).
Fig. 6 shows the evolution of the normalized velocity dispersion and that of
the average eccentricity. Both quantities grow with
$t^{1/2}$ as predicted by the analytical approach in equations (\ref{stf}) and (\ref{etf}).
 The agreements are excellent for $\mu=10^{-7}$ and $10^{-9}$.
Similar to the Brownian motion, the evolution
of the velocity dispersion in an EMS system is a random walk process.
However, the coefficients are not well predicted by the analytic
expression for $\mu =10^{-5}$.   The less satisfactory predictions of
equations (\ref{stf}) and (\ref{etf}) for large masses
may be due to the close separation assumption  $\epsilon  \sim k_0 \mu^{1/3} \ll 1$ being poorly
satisfied in the limit $\mu \ge 10^{-5}$.
We note that in Fig. 6 there are no
very significant transitions in the evolution of $<e>$ when orbit crossing occurs
($\sim 10^3-10^4$ yr according to Fig.3a).  This behavior indicates that
the growth of $<e>$ is a result of a slow diffusion process.

\begin{figure}
 \vspace{4cm}
\includegraphics{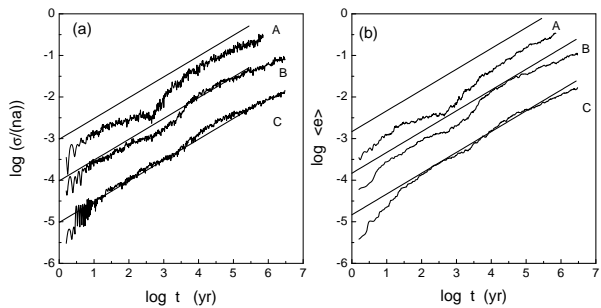}
\caption{\small  Evolution of (a) the variances
of velocity dispersions $\sigma$ normalized by $na$ and (b) the average eccentricities in a
50-planet EMS system with $k_0=5$ and different $\mu$:  A. $\mu=10^{-5}$, B.
$\mu=10^{-7}$, C. $\mu=10^{-9}$. $n,a$ are the mean motion and semi-major axis of each planet.
 The solid lines in  (a) and (b) are
obtained from the analytical formulas (\ref{stf})  and (\ref{etf}),
respectively.  \label{fig6}}
\end{figure}

We now justify the assumption of stochastic phase space. For this task, we calculate the
Lyapunov exponents (LE) at a finite time $\chi(t)$ for the EMS systems.
As is well established for two-planet systems, there is a
well-defined boundary between the regular and chaotic motions which
is demarcated by $k_0 \sim 2 \mu^{2/7}$(Wisdom 1980, Gladman 1993).
%The scaling $\mu^{2/7}$
%was first found by  Wisdom (1980) in a restricted three-body model with a
%resonance overlap criterion.
%Wisdom (1980) found that
%a 2-planet-plus-the-star system with equal planetary masses will be
%chaotic as long as separation of the  two planets $< 2 \mu^{2/7}$,
%where $\mu$ is the mass ratio of each planet to the star.
%Thus the  Hill unstable region is  smaller than the chaotic region
% provided $\mu_1=\mu_2=\mu \le 2\times 10^{-4}$.
However, in EMS systems with $n \ge 3$, $\chi(t)$ may undergo transitions
to a finite value after a long period of time.  The reason for this behavior
is due to the increase of velocity dispersion ($\sim t^{1/2}$) through orbital diffusion.
Orbits initially in a regular  region will finally, though after a very long time,
 become chaotic due to the increase of velocity dispersion.
 Thus we believe the changing from
chaotic motion to regular motion along $k_0$ space is gradual,
and there is no clear boundary between the domains of regular
and chaotic motions (Fig.7). We will discuss this problem elsewhere
(Zhou \& Sun 2007).
In Fig.  8, we map out the Lyapunov time
($T_L$, inverse of LE) as a function of $(k_0, \mu)$.  For computational
simplicity, we consider here only those systems on circular orbits initially.
The chaotic nature of the entire parameter domain calculated justifies our random-phase
assumption.

\begin{figure}
 \vspace{4cm}
\includegraphics{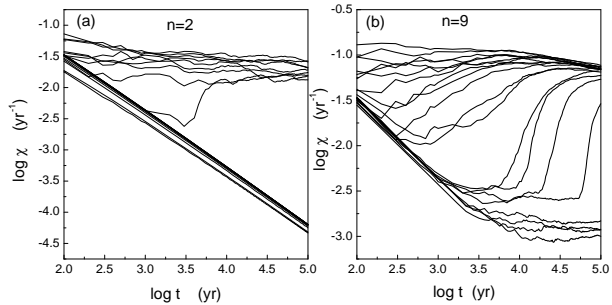}
 \caption{\small  Lyapunov exponents for orbits with
$k_0=2.0+i*0.3, i=0,...,19$ and $\mu=10^{-7}$, $e_0=0$ in  an EMS system with (a)  2 planets,
 (b) 9 planets.  The Lyapunov exponents
are calculated from the variational equations along the solutions. There
are 20 lines in each plot which correspond to i=0,...,19.
The  accumulated value of relative energy error is $\sim 10^{-10}$ for the simulations.
\label{fig7}}
\end{figure}

We also plot in Fig. 8 three lines of constant $T_{\rm c}$ derived from equation (\ref{ff}).
The line corresponds to $T_{\rm c}=10^{4.5}$ yr lies on the boundary between the strongly
(with $T_L < 10^3$ yr) and weakly (with $T_L > 10^3$ yr) chaotic regions. In comparison with
Fig. 4, we find, that the Luapunov time of an EMS system in the strongly
chaotic region is essentially independent of $k_0$,  while in the weakly chaotic
regions, $T_L$ is correlated with $T_{\rm c}$, large $T_{\rm c}$ implies large $T_L$.
This indicates that the Lyapunov time can be either correlated with or independent
of the orbital crossing time, which is a counter example to the conjecture proposed
by Lecar et al. (1992).

\begin{figure}
 \vspace{4cm}
\includegraphics{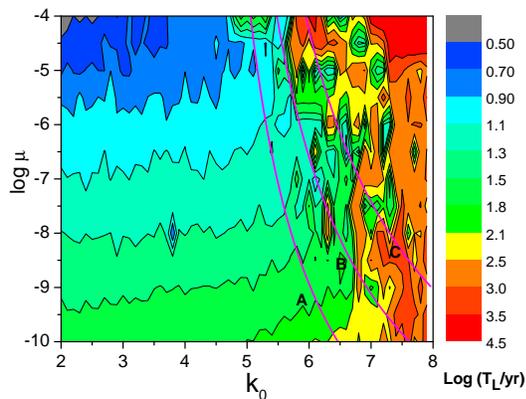}
\caption{\small Lyapunov time, $\log(T_L)$, in the
parameter space $(k_0,\log(\mu))$ of 9-planet EMS systems with $e_0=0$.  The three
dashed lines A,B,C correspond to the crossing time of
$10^4,10^{4.5},10^5$ yr, obtained from equation (\ref{ff}), respectively. \label{fig8}}
\end{figure}

\section{Presence of gas disk}

As indicated in the abstract and introduction, one motivation for
our present study is to consider the growth of protoplanetary embryos
as they undergo a transition from dynamical isolation to post-oligarchic
evolution.  The above analysis on the evolution of EMS systems in a
gas-free environment is appropriate for late stages after
the gas depletion.  In this section, we consider the stability of EMS
systems in a gas environment. Intuitively, gas provides an eccentricity
damping mechanism which may suppress the growth of velocity dispersion
and thus prolong the orbit crossing time.

For illustration, we adopt a fiducial model for the gas surface
density based on the minimum mass nebula model such that
\beq
   \Sigma_g=  \Sigma_0 f_{\rm g} f_{\rm dep}(\frac{a}{\rm 1AU})^{-3/2},
   \label{sg}
\eeq
where  $\Sigma_{0}=2400 {\rm g cm^{-2}}$ and $f_{\rm g}$ is a scaling factor \citep{hay85,il04}. We also use
an idealized prescription to approximate the decline of the gas surface density with
a uniform depletion faction $ f_{\rm dep}=\exp(-t/T_{\rm dep})$.
We adopt a magnitude for the gas depletion time scale to be
$T_{\rm dep} = 3$ Myr based on observations (Haisch et al. 2001).

In a gaseous disk background, a protoplanet with mass ratio $\mu$ suffers a
  gravitational tidal drag, which for simplicity, can be expressed as
 \beq
 {\bf F}_{\rm tidal}=-T_{\rm tidal}^{-1}({\bf V-V_{\rm c}}),
 \eeq
 where  ${\bf V}$ and ${\bf V_{c}}$ are the Keplerian
 and circular velocity of the protoplanet, respectively
 (Kominami \& Ida 2002, Nagasawa et al. 2005).
 The time scale $T_{\rm tidal}$ is  defined as (Ward 1989, Artymowicz 1993)
\beq
 T_{\rm tidal} \approx 0.75 \times 10^{-3} f^{-1}_{\rm g} f^{-1}_{\rm dep}  \mu ^{-1}(\frac{a}{\rm 1AU})^2 ~{\rm yr}.
\label{ttid}
\eeq
For example, the magnitude of $T_{\rm tidal}$ is $\sim 10^4$ yr for a protoplanet with mass ratio $\mu=10^{-7}$.

In principle, an imbalance between the tidal force on either side of the protoplanet's
orbit can lead to ``type I'' migration (Goldreich \& Tremaine 1980, Ward 1997).  But the efficiency
of this process may be suppressed by turbulence and nonlinear response in the disks
%\cite{kll03,lsa04,np04}.
(Koller et al. 2003; Laughlin et al. 2004;
Nelson  \& Papaloizou 2004).
We neglect the effect of type I migration.
However, under the tidal force, eccentricity and inclination damping can also lead to
semi-major axes evolution. To the leading orders of $e$ and $i$ we have,
\beq
\begin{array}{ll}
\frac{1}{a}<\frac{da}{dt}> & = -\frac{1}{8 T_{\rm tidal}}(5e^2+2 i^2),   \\
\frac{1}{e}<\frac{de}{dt}> & =\frac{2}{i}<\frac{di}{dt}>=
-\frac{1}{T_{\rm tidal}}.  \\
\end{array}
\label{dae}
\eeq

The relative importance of eccentricity excitation  by planetary perturbations
versus tidal damping can be estimated by comparing $T_{\rm c}$ with
$T_{\rm tidal}$.  As the damping process proceeds in an exponential fashion,  the
growth of eccentricity is through diffusion,
which does not have a distinct characteristic time scale itself. However,
it has a relevant time scale of $T_{\rm c}$ when orbital crossing is reached.
In addition, $T_{\rm tidal} \propto \Sigma_g^{-1}$.  During gas depletion, $T_{\rm tidal}$ increases as
$f_{\rm dep}$ vanishes and the efficiency of tidal damping weakens.  On general grounds, we anticipate
several possible limiting outcomes:
\begin{description}
  \item[(i)] For closely-separated protoplanets, planetary perturbations are more effective than
  tidal damping, so we expect $T_{\rm c}\ll T_{\rm tidal}$, and orbital crossing occurring
   before the disk is depleted.
  \item[(ii)] In the range of modest separation, the protoplanets' eccentricities
  excited by their mutual interactions are effectively damped by the disk gas.
  Orbital crossing occurs only after severe gas depletion such that $T_{\rm c}\ge  T_{\rm dep}$.
  \item[(iii)] Due to its very long excitation time scale even without a gas background,
  the eccentricities of widely separated protoplanets cannot be excited before the gas
  is severely depleted. Thus $ T_{\rm c}$ is unaffected by the tidal damping.
\end{description}

\begin{figure}
 \vspace{4cm}
\includegraphics{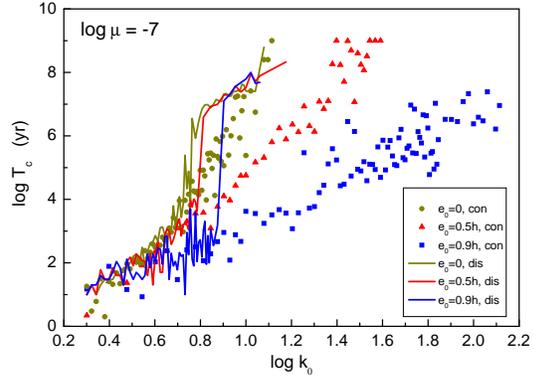}
 \caption{\small Variations of the orbit-crossing
time $T_{\rm c}$ with initial orbital separation  $k_0$ in the 9-planet EMS systems
with a gas-free environment (dots, denoted by `con') or a gas-rich
environment  (curves, denoted by `dis'). Three sets of initial eccentricities are plotted
for both cases.  $h$ is the relative separation defined in equation (\ref{eq2}). \label{fig9}}
\end{figure}

In order to verify these conjectures, we carry out a new set of numerical calculations,
taking into account the tidal dissipation effect.  We adopt a representative value
$\mu = 10^{-7}$.  In Fig. 9, we compare the results of these calculations with
those obtained for EMS systems without any gas.

In systems with $\tilde{e}_0 =0$ and $k_0 < 5$, $T_{\rm c}$ is not affected by the presence
of the disk gas. According to the above classification, we consider these systems as closely
separated.  However, the presence of gas disk delays the crossing time of planets with
modest separation (e.g., $5 \leq  k_0 \leq 8$ in the case of $\tilde{e}_0 =0$)
until gas depletion.  Widely separated systems (with $k_0 > 8$) are not
affected by the presence of the gas.

To illustrate the dominant effect of tidal drag, we study the evolution
of an EMS system during the depletion of the gas disk.
 In Fig.  4b, we plot the evolutions of periapse distance $q=a(1-e)$,
semi-major axis $a$, apoapse distance $Q=a(1+e)$ of an EMS system with modest separation
($k_0 = 8$ and $\tilde{e}_0 = 0.5$).
 Evidently, the eccentricity growth
occurs only after gas depletion for this system.
  Although the magnitude of $T_{\rm c} \sim 10^4$ yr
in a gas-free environment (Fig. 9 and eq. [{\ref{ff}]), the tidal damping effect prolongs
it to $\sim 10^7$ yr.

During the epoch of oligarchic growth, embryos have similar masses
\begin{equation}
\mu \simeq 2 \pi \Sigma_{\rm d} (a_{i+1} - a_i) a_i / M_\ast,
\end{equation}
where $\Sigma_{\rm d}$ is the surface density of the planetesimals and $M_\ast$ is the
stellar mass.  From equations (\ref{kbe}) and (\ref{eq2}), we obtain
\beq
\mu = {(\frac23)^{1/2}} \left( {2 \pi \Sigma_{\rm d} k_0 a^2 \over M_\ast} \right)^{3/2}.
\label{eq:mu}
\eeq
For illustration, we adopt the surface density of a planetesimal disk as
\beq
\Sigma_{\rm d} = 10f_{\rm d} f_{\rm ice} (\frac{a}{\rm 1 AU})^{-3/2}{\rm g~ cm^{-2}},
\eeq
where $f_{\rm d}$ is a scaling constant relative to that of the minimum mass nebula,
$f_{\rm ice}$ is the volatile ice enhancement factor ($f_{\rm ice}=1$ for $a<2.7$ AU and
$f_{\rm ice}=4.2$  for $a>2.7$ AU).
 Substituting it into
equation (\ref{eq:mu}), we obtain the isolation mass, which depends on $k_0$:
\beq
   M_{\rm iso}=0.51\times 10^{-2}M_\oplus \eta k_0^{3/2},
   \label{isom}
   \eeq
where
\beq
   \eta=(f_{\rm d}f_{\rm ice})^{3/2}(\frac{a}{\rm 1AU})^{3/4}
   (\frac{M_*}{M_\odot})^{-3/2}.
   \label{eta}
   \eeq

\begin{figure}
 \vspace{4cm}
\includegraphics{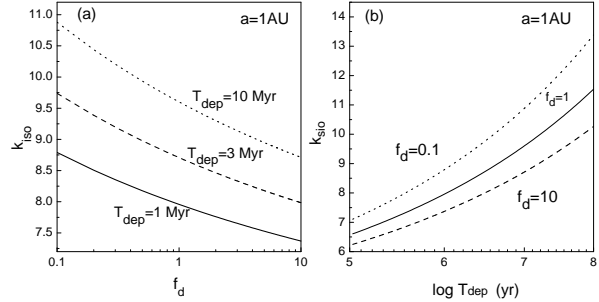}
 \caption{\small Variations of isolation
separation $k_{\rm iso}$ (in the unit of Hill's radius, defined in eq. [2])
with (a) disk enhancement factor $f_{d}$  and (b) gas
depletion time scale $T_{dep}$  at $1AU$.
$k_{\rm iso}$ is calculated from equation (\ref{isok}).
At $T_{dep}=3$Myr and $f_{\rm d}=1$, which corresponds to a surface density
$10 {\rm g ~cm^{-2}}$ of dust at 1AU, the isolation separation
$\approx  8.7 $  Hill's Radius and the isolation
mass  $\approx 0.13M_\oplus$.
 \label{fig10}}
\end{figure}

During the formation of protoplanets, orbital crossing induces protoplanets to undergo cohesive
collisions, mass growth, and increasing separation.  This stage corresponds to case (i).  Prior to the gas
depletion, the value of $k_0$ for an EMS system increases until the perturbation between
protoplanets can no longer dominate their tidal interaction with the disk.
During this  end stage, which corresponds to case (ii), the evolution of $\tilde{e}$, $\mu$, and $k_0$ becomes
stalled in a gas-rich environment.  Until the gas is severely depleted, the
embryos attain an isolation mass, which can be derived from the condition
that $T_{\rm c} \sim  T_{\rm dep}$. Substituting this condition with
$T_{\rm c}$ from equation (\ref{ff}) for circular orbits $(\tilde{e}=0)$, and
using the isolation mass determined from equation (\ref{isom}), we get the critical
separation of an isolation mass:
\beq
     \log(k_{\rm iso})=\sqrt{b^2+0.61c}-b,
     \label{isok}
\eeq
where
\beq
\begin{array}{l}
    b=2.8+0.33\log \eta ,  \\
    c=3.6+0.67\log \eta+\log T_{\rm dep},\\
   \end{array}
\eeq
and $\eta$ is defined in equation (\ref{eta}).
%For application (see below), we use
%eq. [\ref{eq:mu}] to substitute $\mu$ with $k_0$ and then evaluate with
%eq. [\ref{ff}] the magnitude of $k_0$ as a function of $\Sigma_{\rm d}$ and $T_{\rm
%dep}$.
% For illustration purpose, we adopt $\Sigma_{\rm d} = f_{\rm d} \Sigma_{\rm min}$
%where $f_{\rm d}$ is a constant and $\Sigma_{\rm min} = 10 (a/ {\rm AU})^{-3/2}
%{g \ cm}^{-2}$ is that of planetesimals in a minimum mass nebula.
In Fig. 10, we plot $k_{\rm iso}$ as a function of $f_{\rm d}$ and
$T_{\rm dep}$ at 1AU around a solar-type star.  These results
indicate that $k_{\rm iso}$ decreases slightly with the increase of disk mass,
which is consistent qualitatively with the numerical results of
Kokubo and Ida (2002).  The isolation separation
$k_{\rm iso}$ and isolation mass $M_{\rm iso}$
 of the planets are plotted  in the whole disk
region for different $T_{\rm dep}$ (Fig. 11) and  $f_{\rm d}$ (Fig. 12). For $T_{\rm dep}
\simeq 3 \times 10^6$ yr and $f_{\rm d}=1$, the isolation mass of embryos
is $\sim 0.13 M_\oplus$ and their critical separation $k_{\rm iso} \simeq 8.7$.
These results
 support the assumption that isolated embryos are
separated by a distance that is approximately ten times their Hill's
radii \cite{il04}.

\begin{figure}
 \vspace{4cm}
 \includegraphics{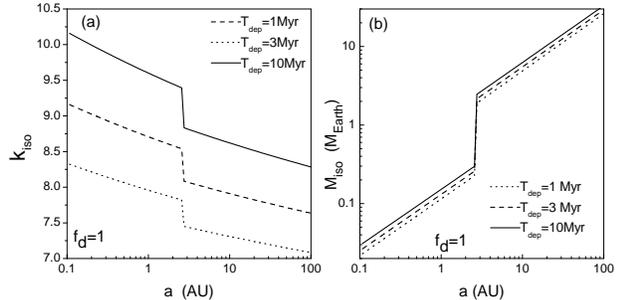}
\caption{\small Variations of (a) isolation separation $k_{\rm iso}$ and
(b) isolation masses $M_{\rm iso}$ with radial distance to the
star for disk enhancement factor $f_{d}=1$ and different gas
depletion time scale $T_{dep}$.
$k_{\rm iso}$ and  $M_{\rm iso}$ are calculated from equations (\ref{isok}) and (\ref{isom}),
respectively.
 \label{fig11}}
\end{figure}

\begin{figure}
 \vspace{3.5cm}
\includegraphics{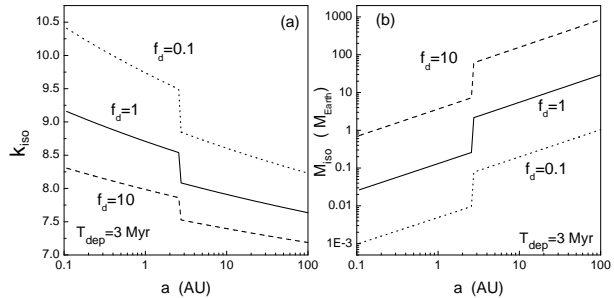} \caption{ Variations of (a) isolation
separation $k_{\rm iso}$ and (b) isolation masses $M_{\rm iso}$
 with radial distance to the star for different disk enhancement factor $f_{\rm d}$.
 $k_{\rm iso}$ and  $M_{\rm iso}$ are calculated from equations (\ref{isok}) and (\ref{isom}), respectively.
 $f_{d}$ is disk enhancement factor
 and $T_{dep}=3 Myr$ is the time scale of gas depletion  .
\label{fig12}}
\end{figure}

\section{Conclusions and applications}

In this paper, we extend the study  on the orbital crossing time ($T_{\rm c}$)  of
 n-planet systems with equal planetary masses
and  separation (EMS systems), which was investigated by Chambers  et al. (1996)
 and Yoshinaga et al. (1999).
We find $T_{\rm c}$  of  EMS systems can be formulated
as a power law in equation (\ref{ff}). The results have the following implications:

(i) The onset of instability  in an EMS system mainly depends on the initial separation ($k_0$).
A qualitative inspection of equation (\ref{ff}) indicates that doubling $k_0$ can enlarge $T_{\rm c}$ by several orders of magnitude.
In two systems with identical $k_0$, $T_{\rm c}$ increases with the planetary masses.
This counter-intuitive result is due to the mass dependence of the planetary Hill's
radii.  For constant $k_0$ values, the un-normalized physical separation
between planets, {\it i.e.} $a_{i+1} - a_i$, increases with their masses.

ii) In a protostellar disk, a large population of low mass
planetesimals emerge quickly.  During the early stage of disk
evolution, the crossing time of planetesimals is relatively short.
So the planetesimals will collide, merge  and grow,
leading to the decline of their number density.
Equation (\ref{eq:mu}) suggests that  $k_0$ of embryos increases with
$\mu$.  Since $T_{\rm c}$ increases rapidly with
$k_0$, the eccentricity growth due to dynamical diffusion is
slowed down. In a gas-rich environment, the eccentricities of embryos are
also damped by their interaction with the disk gas.  With mass
distribution comparable to that of the minimum mass nebula, tidal
damping becomes effective when embryos merge into bodies separated
by $k_0 > 5$.  As the  orbits of embryos are circularized, their
growth is stalled. This result is supported by the simulations of
planetesimal growth in a minimum mass environment, which leads to
embryos with asymptotic masses of $\sim 10^{25}$ g on nearly circular
orbits with separation $\sim 10$ times of their Hill's radii \citep{ki98}.

\begin{figure}
 \vspace{3cm}
\includegraphics{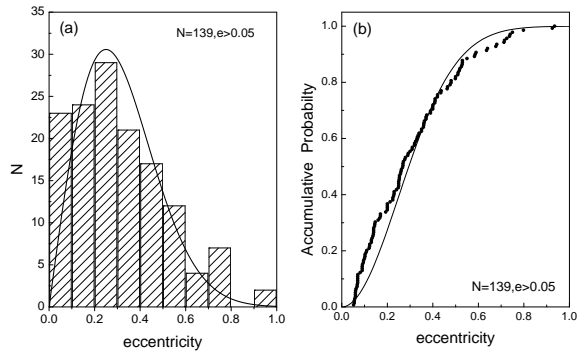}
\caption{\small Eccentricity distribution of the 139
observed extra solar planets with eccentricities $>0.05$ (from the
data of Butler et al. 2006). The average eccentricity of these 139
planets is $<e>=0.31$. (a) The histogram of the distribution in
eccentricity.  The solid line is the fit of a Rayleigh
distribution by equation  (\ref{ray})
 with $\sigma=0.25$ and a scaling factor of $12.6$.
 (b) The corresponding accumulative  distributions
 for the observed  139 planets with $e>0.5$ (dotted line)
 and for the best-fit Rayleigh distribution (solid line).
 %we take $\sigma=<e>/\sqrt{\pi/2}=0.25$
\label{fig13}}
\end{figure}

iii) The gas accretion rate from protostellar disks onto their
central stars decreases exponentially on a characteristic time
scale of $\sim 3 \times 10^6$ yr \cite{hartmann2}.
Presumably the magnitude of $\Sigma_g$ also decreases on a similar
time scale, hence the tidal damping would become
less effective. Subsequently, dynamical equilibria (in which
$T_{\rm c} \sim T_{\rm tidal}$) are maintained with increasing
separation, $k_0$, while embryos merge, grow, and space out,
albeit at a much slower pace. When the disk gas is severely
depleted within a few depletion time scales, $T_{\rm tidal}$
becomes large compared with $T_{\rm dep}$ and the embryo-disk
interaction is no longer effective. In a disk with minimum mass nebula ($f_{\rm d}=1$),
the isolation separation ($k_{\rm iso}$) and isolation mass ($M_{\rm iso}$)  of embryo
determined by $T_{\rm c} \sim T_{\rm dep}$ are $8.7~R_H$ and $0.13 ~M_\oplus$ at 1 AU, respectively,
while at 5 AU, $k_{\rm iso}=8.0 R_H$, $M_{\rm iso} = 3.3~M_\oplus$.
 In a following paper, we will apply these results to evaluate whether
embryos can attain several earth masses while there is adequate
residual gas supply in the disk for them to acquire their gaseous
envelopes and grow into gas giants.

iv) In the radial velocity surveys, no planet is detected in a majority of
the target stars.  The failure for the emergence of any gas giant planets
does not prevent the embryos to grow after the total gas depletion. The eccentricity
of the residual embryos
increases through a post-oligarchic random walk process.  As the orbital crossing leads to
giant impacts, mass growth, and widening separation, $T_{\rm c}$ increases
until it is comparable to the age of the system.  Since $T_{\rm c}$ is a steeply
increasing function of $k_0$, the  separation of embryos is unlikely to
exceed $10 R_H$ by much.

v) However, around stars with known gas giant planets, the gas depletion may lead
to a sweeping secular resonance which has the potential to shake up the
kinematic structure of the ``isolated embryos''.  In Fig. 3b we show that
for EMS systems which ended up with $k_0 > 10-12$, $T_{\rm c}$ exceeds the age of the Solar
System.  Indeed, the actual value of $k_0$ is in this range, which accounts
for the dynamical stability of the Solar System.

vi) A significant fraction of stars with known planets show signs of
additional planets. Such systems generally have eccentricities much
larger than those of most planets in the Solar System.
The emergence of the first-born gas giants induces the gap formation
in their nascent disks and the accumulation of planetesimals
exterior to the outer edge of the gap (Bryden  et al. 1999).  This
process promotes the formation of multiple-planet systems.  In contrast
to the embryos, the spacing between the gas giants may be regulated
by various migration processes and their masses are determining by the disks'
thickness-to-radius ratio.

Modest ranges of $k_0$ and $\mu$ values are anticipated when a system with giant planets
forms.
Gas giants emerging too closely ($k_0 < 5$) will undergo
orbital crossing (Fig. 3b), close encounters, and cohesive collisions.
Gas giants formed with $\mu \sim 10^{-3}$ and $k_0 \sim 5.5$ have $T_{\rm c}
\sim T_{\rm dep}$ whereas those with $k_0 \sim 6$ have $T_{\rm c} \sim 1$ Gyr.
The discussion under item iii) suggests that close encounters and mergers
may occur among these gas giant planets, which may provide a mechanism
for generating the large observed eccentricities.  We expect a
considerable dispersion in diffusion rate and the asymptotic eccentricities
of these systems,  because gap formation may reduce the efficiency of
eccentricity damping  by the planet-disk tidal interaction.  Close
encounters between planets with relative large masses $\mu \sim 10^{-3}$
can also lead to nonlinear effects such as changes of semi-major axis.  For
gas giants formed with $k_0 > 6$, neither tidal damping nor mutual
perturbations of planets are effective and they are likely to retain their original
low-eccentricity orbits.

vii) We speculate that the large observed eccentricities among the extra
solar planets may be due to scattering between multiple
 planets. In
\S3, we show that the asymptotic eccentricities of the planets
have a Rayleigh distribution, similar to the case of planetesimal growth
(Ida \& Makino 1992, Palmer et al. 1993, Lissauer \& Stewart 1993).
In Fig. 13, the eccentricity distribution of
the observed extra solar planets is fit by a Rayleigh distribution.
The close agreement provides  evidence that the eccentricity
of extra solar planets may be excited by the inter-planetary scattering\footnote{We notice after we finished the manuscript that,
 a similar conclusion is also obtained in a recent work
 by Mario \& Scott (2007).}.

%We will carry out detailed numerical simulation to verify this conjecture.
%The contribution of two gas giants, Jupiter and Saturn,
%dominates the gravity of the planets in the solar system.  Neglecting all other
%contribution, this $n=2$ EMS system is Hill stable since its $k_0 > 3.5$.  In
%principle, their orbits will never cross.  However, the contribution of other
%planets leads to dynamical diffusion which promotes their average eccentricity
%to increase $\propto t^{1/2}$.  The diffusion rate is
 %   Thus for the case $k_0>3.5$ when the two cores will never cross in the $n=2$  case,
 %  the presence of additional cores induce the merge or ejection of
 %  the former two cores. Also the scaling of $1/2$ in eccentricity evolution
 %  is different with that of the planetesimal evolution
  %  by viscous stirring, where a scaling of $1/4$ is found numerically (Ida \& Makino 1992)

\acknowledgments

 We thank the anonymous referee for valuable suggestions, and
 Dr. S. Aarseth for improving the manuscript.
 This work is supported  by %Natural Science Foundation of China
NSFC(10233020,10778603), NCET (04-0468),
NASA (NAGS5-11779, NNG04G-191G, NNG06-GH45G), JPL (1270927), NSF(AST-0507424, PHY99-0794).

\end{document}